\documentclass[aps,twocolumn,showpacs,preprintnumbers,amsmath,amssymb]{revtex4}

\usepackage{graphicx}% Include figure files
\usepackage{dcolumn}% Align table columns on decimal point
\usepackage{bm}% bold math

\newcommand{\ket}[1]{\left |\mbox{$#1$}\right\rangle}
\newcommand{\bra}[1]{\left\langle\mbox{$#1$}\right |}

\begin{document}

\title{Theoretical analysis of the optimal conditions \\
for photon-spin quantum state transfer
}
\author{Y. Rikitake$^{1,2}$, H. Imamura$^{1,2}$ and  H. Kosaka$^{1,3}$}
\affiliation{
$^1$CREST-JST, 4-1-8 Honcho, Kawaguchi, Saitama 322-0012, Japan
\\
$^2$Nanotechnology Research Institute, National Institute of
Advanced Industrial Science and Technology, 1-1-1 Umezono, Tsukuba,
Ibaraki 305-8568, Japan
\\
$^3$Research Institute of Electrical Communication, Tohoku University,
Sendai 980-8577, Japan
}

%[NOTE. PLEASE REMEMBER TO ADD THE APPROPRIATE PACS NUMBERS]

%%%%%%%%%%%%%%%%%%%%%%%%%%%%%%%%%%%%%%%%%%%%%%%%%%
%%%%% abstract
%%%%%%%%%%%%%%%%%%%%%%%%%%%%%%%%%%%%%%%%%%%%%%%%%
\begin{abstract}
 We analyzed the yield and fidelity of the
 quantum state transfer (QST) from a photon polarization qubit
 to an electron spin qubit in a spin-coherent photo detector
 consisting of a semiconductor quantum dot.
 We used a model consisting of the quantum dot, where the QST is
 carried out, coupled with a photonic cavity.
 We determined the optimal conditions that allow the realization of
 both high-yield and high-fidelity QST.
\end{abstract}

\maketitle

\begin{figure}[t]
 \begin{center}
  \begin{tabular}{ll}
    (a) & (b) \\
   \includegraphics[width=0.4\linewidth]{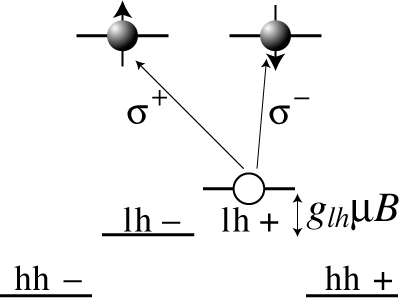}
   &
   \includegraphics[width=0.4\linewidth]{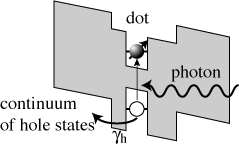}
  \end{tabular}
 \end{center}
 \caption{(a) Energy levels of the quantum dot.
 From the $\ket{lh+}$ state, an electron with the $\ket{\uparrow}$
 ($\ket{\downarrow}$) spin state is optically excited by the
 right-handed (left-handed) circularly polarized photon $\ket{\sigma^{+}}$
 ($\ket{\sigma^{-}}$).
 (b) Schematic of band diagram of the quantum dot.
 The continuum of the hole states is attached to the quantum dot
 via the tunneling barrier.
 The created hole in the dot is extracted to this continuum
 with the extraction rate $\gamma_h$.
 } 
 \label{fig:SelectionRule}
 \label{fig:energyDiagram}
\end{figure}

A quantum repeater is a promising
technology \cite{briegel98,childress05,taylor05}
which makes it possible to drastically expand the distance of quantum 
key distribution as well as to realize scalable quantum networks.
The quantum repeater requires
not only messenger qubits but also processing qubits.
A photon is, of course, the most convenient candidate for 
the messenger qubit \cite{gisin02,gobby04};
however, it is not suitable to be used as a processing qubit
because quantum information storage and two-qubit quantum
operations are very difficult to perform.
On the other hand, an electron spin in a semiconductor quantum dot is one 
of the most convenient candidates for processing qubits
since one- and two-bit quantum gate operations
can now be performed through the application of electric and/or
magnetic fields
\cite{loss98,taylor03,petta05,petta05a}.
Therefore, 
it is important to investigate the possibility of
a photon-spin quantum state transfer (QST) 
\cite{vrijen01,duan01,kosaka03,muto05}
which would transfer quantum information from a 
photon-polarization (photon qubit) to an electron-spin
(spin qubit), as an interface device for quantum repeaters.
Such a photon-spin QST can be performed 
using a semiconductor spin-coherent photo detector,
as proposed by Vrijen and Yablonovitch \cite{vrijen01}, who    
showed that the well-known optical orientation
in a semiconductor heterostructure can be used for QST
with the help of $g$-factor engineering
\cite{kiselev98,ivchenko98,matveev00,kosaka01,salis01,salis03,nitta03,lin04,nitta04}
and strain engineering \cite{lin91,nakaoka04,nakaoka05}.
The photo detector has a quantum dot whose energy levels are shown
in Fig. \ref{fig:SelectionRule}(a).
The $g$-factor of the electron spin is tuned to zero 
($g_e=0$).
By using strain engineering and applying a magnetic field,
we can arrange the light-hole state 
$\ket{lh+}=\ket{3/2,1/3}+\ket{3/2,-1/2}$ 
to the topmost level of the valence band.
According to the selection rule \cite{vrijen01}, 
a spin-up (down) electron $\ket{\uparrow}$ ($\ket{\downarrow}$)
is excited by a right-handed (left-handed) circularly polarized photon
$\ket{\sigma^{+}}$ ($\ket{\sigma^{-}}$) from the $\ket{lh+}$ state.
%% add when resubmit
As shown in Fig. \ref{fig:SelectionRule}(b),
the quantum dot is connected to the continuum of the hole state
and the created hole extracted from the dot to the continuum.
By eliminating the hole in the $\ket{lh+}$ state,
we complete the QST from the photon qubit
$\alpha_{+}\ket{\sigma^{+}}+\alpha_{-}\ket{\sigma^{-}}$
to the spin qubit
$\alpha_{+}\ket{\uparrow}+\alpha_{-}\ket{\downarrow}$.

One of the obstacles to high-fidelity QST in
a spin-coherent semiconductor photo detector is 
the exchange interaction between the electron 
and the simultaneously created hole.
The electron-hole exchange interaction disturbs
the electron spin state and decreases fidelity.
In our previous work \cite{rikitake06}, 
we proposed the use of resonant tunneling 
in a double-well structure \cite{gurvitz91, cohen93}
in order to avoid the electron-hole exchange interaction.
However, we considered
the QST process only after the creation of the exciton,
and did not take into account the exciton creation process itself; 
thus, our previous work lacked discussion about the efficiency, or 
yield of the QST.
In the present study, we analyzed the QST process 
from the photon qubit to the spin qubit
in a spin-coherent semiconductor photo detector,
considering the whole QST process from the input photon
to the generated electron spin via the dot exciton state.
We present here the conditions necessary to produce
a high-yield, high-fidelity QST.

%----- fig:SelectionRule

%\section{Model}

%%%%%%%%%%%%%%%%%%%%%%%%%%%%%%%%%%%%%%%%%%%%%%%%%%
%%%%% model
%%%%%%%%%%%%%%%%%%%%%%%%%%%%%%%%%%%%%%%%%%%%%%%%%

\begin{figure}[t]
 \begin{center}
   \includegraphics[width=0.6\linewidth,clip]{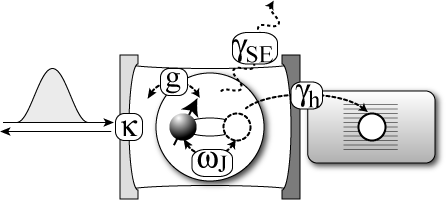}
 \end{center}
 \caption{The theoretical model considered.
 The model includes the input/output photon field,
 the cavity, and the quantum dot where the QST is carried out.
 This model is characterized by the following parameters:
 the cavity damping rate $\kappa$, the cavity-dot interaction $g$,
 the extraction rate of the created hole $\gamma_h$,
 the electron-hole exchange interaction in the dot $\omega_J$,
 and the spontaneous emission rate $\gamma_h$.}
 \label{fig:Model}
\end{figure}

We examined the model shown in Fig. \ref{fig:Model}.
The quantum dot is located in the photonic cavity.
The quantum dot is also connected to the continuum of the hole
through the tunneling barrier in order to extract the created hole.
%%%
The input photon propagates the $x$-axis
and enters the cavity (one-sided cavity)\cite{kojima03,kojima07,koshino04b}.
The cavity photon excites the electron-hole pair
or exciton in the dot.
By applying appropriate selection rule,
the electron spin state is determined by
the input photon polarization state.
Once the electron-hole pair is excited in the quantum dot,
the created hole is extracted to the continuum
and the electron spin is left in the dot.
The QST process is then completed.
In addition to the above processes,
we also considered the electron-hole exchange interaction
and the spontaneous emission from the dot.
The electron-hole exchange interaction
modifies the orientation of the electron-spin, while
the hole is in the dot and therefore reduces 
the fidelity of the QST.
Spontaneous emission leads to a decrease in yield.

%%%%% state vector
The state vector of the system under consideration is expressed as
\begin{align}
 \ket{\Psi(t)}
 & =
 \sum_{\sigma=\pm}\int dx 
 \phi^{\text{ph}}_{\sigma}(x;t)\ket{x,\sigma}
 +
 \sum_{\sigma=\pm}
 \phi^{\text{cav}}_{\sigma}(t)\ket{\text{cav},\sigma}
 \nonumber
 \\
 & +
 \sum_{s=\uparrow\downarrow}
 \varphi_{s}^{\text{ex}}(t)\ket{\text{ex},s}
 +
 \sum_{\mu}
 \phi^{\text{noncav}}_{\mu}(t)\ket{\mu}
 \nonumber
 \\
 &  +
 \sum_{s=\uparrow\downarrow}\sum_{l}
 \varphi^{\text{spin}}_{ls}(t)\ket{l,s},
 \label{eq:Psi}
\end{align}
where $\ket{x,\sigma}$ denotes the input/output photon state at
position $x$ and polarization $\sigma=\sigma^{\pm}$;
$\ket{\text{cav},\sigma}$ the cavity photon with polarization $\sigma$;
$\ket{\text{ex},s}$ the exciton state in the dot with electron 
spin state $s=\uparrow,\downarrow$;
$\ket{\mu}$ the external noncavity photon ($\mu$ represents
the wave number and corresponding polarization);
and $\ket{l,s}$ the state in which the electron with spin state $s$
is in the dot and the hole is in the continuum state $l$.
Here $\phi^{\text{ph}}_{\sigma}(x;t)$,
$\phi^{\text{cav}}_{\sigma}(t)$, $\varphi_{s}^{\text{ex}}(t)$,
$ \varphi^{\text{spin}}_{ls}(t)$, and $\phi^{\text{noncav}}_{\mu}(t)$
are coefficients to be determined by 
solving the Schr\"{o}dinger equation.
For later use, we name the state represented by 
the last term in Eq. \eqref{eq:Psi} the ``spin state''.

%----- fig:Model

%%%%% Hamiltonian
The Hamiltonian of the system is given by
\begin{align}
 H 
 & = 
 H_{\text{ph}} + H_{\text{ph-cav}} +H_{\text{cav}}
 + H_{\text{cav-ex}} + H_{\text{ex}}
 \nonumber
 \\
 & +
 H_{\text{SE}} + H_{\text{noncav}}
 + H_T + H_{\text{spin}},
\end{align}
each of whose terms are defined as follows, 
under the condition $\hbar=1$:
\begin{align}
 %%%%
 H_{\text{ph}}
 & =
 \sum_{\sigma=\sigma^{\pm}}\sum_{k}
 c k\ket{k,\sigma}\bra{k,\sigma},
 \label{eq:H_ph}
 \\
 %%%%
 H_{\text{ph-cav}}
 & =
 \sum_{\sigma=\sigma{\pm}}\sum_{k}
 \{
 \kappa_k \ket{\text{cav},\sigma}\bra{k,\sigma}
 + \text{h.c.}
 \},
 \label{eq:H_ph-cav}
 \\
 %%%%
 H_{\text{cav}}
 & =
 \sum_{\sigma=\sigma{\pm}}
 \omega_{\text{cav}}\ket{\text{cav},\sigma}\bra{\text{cav},\sigma},
 \label{eq:H_cav}
 \\
 %%%%
 H_{\text{cav-ex}}
 & =
 \sum_{\sigma=\pm}
 \{
 g\ket{\text{ex},s(\sigma)}\bra{\text{cav},\sigma}
 + \text{h.c.}
 \},
 \label{eq:H_cav-ex}
 \\
 %%%%
 H_{\text{ex}}
 & =
 \sum_{s=\uparrow\downarrow}
 \omega_{\text{ex}} \ket{\text{ex},s}\bra{\text{ex},s}
 +
 \omega_{J} \ket{\text{ex},s}\bra{\text{ex},\bar{s}},
 \label{eq:H_ex}
 \\
 %%%%
 H_{\text{SE}}
 & =
 \sum_{s=\uparrow\downarrow}\sum_{\mu}
 \{
 \gamma_{s\mu}\ket{\mu}\bra{\text{ex},s}
 + \text{h.c.}
 \}
 \label{eq:H_SE}
 \\
 %%%%
 H_{\text{noncav}}
 & =
 \sum_{\mu}
 \omega_{\mu}\ket{\mu}\bra{\mu},
 \label{eq:H_noncav}
 \\
 %%%%
 H_T
 & =
 \sum_{s=\uparrow\downarrow}\sum_{l}
 \{
 W_{l}\ket{l,s}\bra{\text{ex},s}
 + \text{h.c.}
 \},
 \label{eq:H_T}
 \\
 %%%%
 H_{\text{spin}}
 & =
 \sum_{s=\uparrow\downarrow}\sum_{l}
 (\omega_{e}+\omega_{l})\ket{l,s}\bra{l,s}.
 \label{eq:H_spin}
\end{align}
The Hamiltonian for the input/output photon is
$H_{\text{ph}}$, 
where $k$ is wave number of the input/output photon, and
$c$ is the speed of light.
Note that we use the wave number representation $k$ instead of 
the position representation $x$ used in Eq. \eqref{eq:Psi}.
The interaction between the input/output photon and
the cavity photon is represented by $H_{\text{ph-cav}}$.
With respect to the coupling constants $\kappa_k$ in $H_{\text{ph-cav}}$,
we use the flat-band assumption $\kappa_k = \sqrt{{c\kappa}/{\pi}}$
\cite{koshino04b},
where $\kappa$ is the cavity damping rate.
The Hamiltonian for the cavity photon is $H_{\text{cav}}$,
and $\omega_{\text{cav}}$ denotes the energy of the cavity mode photon.
The cavity photon excites the dot exciton; this excitation is
represented by $H_{\text{cav-ex}}$, which is characterized 
by the cavity-dot coupling $g$.
The index of the photon polarization $\sigma(s)$ 
represents the selection rule:
$\sigma(\uparrow)=\sigma^{+}, \sigma(\downarrow)=\sigma^{-}$.
The Hamiltonian for the dot exciton is $H_{\text{ex}}$, 
$\omega_{\text{ex}}$ is the excitation energy, 
$\omega_J$ is the electron-hole exchange interaction 
which flips the electron spin state, and
$\bar{s}$ represents the opposite spin to $s$.
The spontaneous emission process from the dot exciton is represented
as $H_{\text{SE}}$.
The emitted photon escapes to the external noncavity photon mode
$\mu$ with energy $\omega_{\mu}$ and $H_{\text{noncav}}$
is the Hamiltonian of the noncavity photon.
The extraction of the created hole in the dot
is represented by the hole's tunneling Hamiltonian $H_T$,
where $W_l$ is the tunneling matrix elements.
The Hamiltonian for the spin state 
where the electron spin is left in the dot
and the hole is extracted to the continuum is $H_{\text{spin}}$.
The energy of the electron in the dot is $\omega_e$, and 
that of the hole in the continuum state $l$ is $\omega_l$.

%%%%%%%%%%%%%%%%%%%%%%%%%%%%%%%%%%%%%%%%%%%%%%%%%%
%%%%% analysis
%%%%%%%%%%%%%%%%%%%%%%%%%%%%%%%%%%%%%%%%%%%%%%%%%

In order to analyze the QST process in this model,
we must consider the scattering problem from 
the initial input photon state to the final spin state.
The initial state at $t=0$ is written as
\begin{align}
 \ket{\Psi(0)} 
 = 
 \int dx
 \phi^{\text{ph}}(x;t=0) 
 \sum_{\sigma=\sigma^{\pm}}\alpha_{\sigma}\ket{x,\sigma}, 
 \label{eq:init_phi}
\end{align}
where $\alpha_{+}$ and $\alpha_{-}$ are the probability amplitudes 
which characterize the superposition state of the photon qubit
$\alpha_+\ket{\sigma_+}+\alpha_{-}\ket{\sigma_-}$.
The coefficient $\phi^{\text{ph}}(x;t=0)$ represents 
the input photon wave packet.
We consider a Gaussian wave packet defined as
\begin{align}
 \phi^{\text{ph}}(x;0)
 =
 \dfrac{e^{-i\omega_{\text{ph}} x/c}
 e^{-\Delta\omega_{\text{ph}}^2(x-x_{0})^2/2c^2}}
 {\pi^{1/4}\sqrt{c/\Delta\omega_{\text{ph}}}},
 \label{eq:init_Psi}
\end{align}
where $x_0$ ($<0$) is the center position of the wave packet 
at $t=0$,
$\omega_{\text{ph}}$ the center frequency of the photon and
$\Delta\omega_{\text{ph}}$ the bandwidth of the input photon.

The extraction process of the hole 
is characterized by the spectral function
defined by $\gamma_h(\omega)=\pi\sum_l |W_l|^2 \delta(\omega-\omega_l)$.
In the present analysis, we treat this spectral function
$\gamma_h(\omega)$ as the constant $\gamma_h$.
This treatment corresponds to the Markov approximation with respect
to the extraction process, and
the constant $\gamma_{h}$ represents the extraction rate of the hole
from the dot to the continuum.
We also apply the Markov approximation to the 
spontaneous emission process; that is,
we treat the spectral function
$\gamma_{\text{SE}}(\omega) = \pi\sum_{\mu} |\gamma_{s\mu}|^2
\delta(\omega-\omega_{\mu})$ as the constant $\gamma_{\text{SE}}$,
where $\gamma_{\text{SE}}$ is the spontaneous emission rate of 
the dot exciton.

By employing the Schr\"{o}dinger equation and 
the above Markov approximations,
we obtain the equations for the probability amplitudes
of the cavity photon and the dot exciton as,
\begin{align}
 %%%% 
 \dfrac{d\phi^{\text{cav}}_{\sigma}}{dt}(t)
 & =
 (-i\omega_{\text{cav}}-\kappa)\phi^{\text{cav}}_{\sigma}(t)
 -ig^{\ast}\varphi^{\text{ex}}_{s(\sigma)}(t)
 \nonumber
 \\
 & -i\sqrt{2c\kappa}\alpha_{\sigma}\phi^{\text{ph}}(-ct;0),
 \label{eq:dot_phi_cav}
 \\
 %%%%
 \dfrac{d\varphi^{\text{ex}}_{s}}{dt}(t)
 & =
 (-i\omega_{\text{ex}}-\gamma_h-\gamma_{\text{SE}})\varphi^{\text{ex}}_{s}(t)
 -i\omega_J\varphi^{\text{ex}}_{\bar{s}}(t)
 \nonumber
 \\
 & -ig\phi^{\text{cav}}_{\sigma(s)}(t).
 \label{eq:dot_varphi_ex}
\end{align}
By setting the input photon polarization $\alpha_{\sigma}$
and the shape of the wave packet $\phi^{\text{ph}}(x=-ct;0)$ as
Eq. \eqref{eq:init_Psi}, we can solve these equations and
obtain the time evolutions of $\phi^{\text{cav}}_{\sigma}(t)$ 
and $\varphi^{\text{ex}}_{s}(t)$.
The probability amplitude of the spin state 
$\varphi^{\text{spin}}_{ls}(t)$ is obtained from 
that of the exciton state $\varphi^{\text{ex}}_{s}(t)$ as
\begin{align}
 \varphi^{\text{spin}}_{ls}(t)
 & =
 -iW_l\int_0^{t}dt^{\prime}
 e^{-i(\omega_e+\omega_l)(t-t^{\prime})}
 \varphi^{\text{ex}}_{s}(t^{\prime}).
\end{align}
Let us consider the final state of the system for $t\to\infty$.
The final electron spin state is
characterized by the reduced density matrix $\rho^{\text{spin}}$
defined as
\begin{align}
 \rho^{\text{spin}}_{ss^{\prime}}
 & =
 \sum_l
 \varphi^{\text{spin}}_{ls}(\infty)
 \varphi^{\text{spin}}_{ls^{\prime}}(\infty)^{\ast}
\end{align}
for $s,s^{\prime}=\uparrow,\downarrow$.
The efficiency of the QST is represented by the yield defined as 
$P=\rho^{\text{spin}}_{\uparrow\uparrow}+\rho^{\text{spin}}_{\downarrow\downarrow}$,
which represents the probability that the electron spin is 
left in the dot at the final state.
Using $\rho_{\text{spin}}$, 
we can define the fidelity of the QST as 
$F=\bra{\Psi_{I}}\rho^{\text{spin}}\ket{\Psi_{I}}/P$,
where
$\ket{\Psi_{I}} = \alpha_{+}\ket{\uparrow} + \alpha_{-}\ket{\downarrow}$
is the ideal spin state.
Here we note that the fidelity is normalized by yield $P$;
that is, we take into account the fidelity only of the generated spin
at the final state.
Straightforward calculations then give us the expressions for
the yield and fidelity as
\begin{align}
 P & = \sum_{\nu=\pm} |\beta_{\nu}|^2 I_{\nu\nu}, \text{and}
 \\ 
 F & = (\sum_{\nu=\pm} |\beta_{\nu}|^4 I_{\nu\nu}
 + 2|\beta_{+}\beta_{-}|^2  I_{+-})/P,
\end{align}
respectively, where $\beta_{\pm}=(\alpha_{+}\pm\alpha_{-})/\sqrt{2}$.
The coefficients $I_{\nu\nu^{\prime}}$ ($\nu,\nu^{\prime}=\pm$) 
are defined as follows:
\begin{align}
 I_{\nu\nu^{\prime}}
 & =
 \dfrac{4}{\pi}\dfrac{\kappa|g|^2\gamma_h}{\Delta\omega_{\text{ph}}}
 \int d\omega
 \dfrac{\exp[-(\omega-\omega_{\text{ph}}+\omega_{\text{ex}})^2
 /\Delta\omega_{\text{ph}}^2]}
 {D_{\nu}(\omega)D_{\nu^{\prime}}(\omega)^{\ast}},
\end{align}
where
\begin{align}
 D_{\nu}(\omega)
 & =
 \{i(\omega+\omega_{\text{cav}}-\omega_{\text{ex}})+\kappa\}
 \{i(\omega+\nu\omega_J)-\gamma_h-\gamma_{\text{SE}}\}
 \nonumber
 \\
 & +
 |g|^2.
\end{align}

%%%% Fixed parameters
The parameters we used in the present paper are as follows:
input photon bandwidth $\Delta\omega_{\text{ph}}=5\text{GHz}$, 
electron-hole exchange interaction $\omega_J=5\text{GHz}$,
and spontaneous emission rate $\gamma_{\text{SE}}=1\text{GHz}$.
We used the typical values in a GaAs quantum dot
for $\gamma_{\text{SE}}$ and $\omega_J$ \cite{gammon96,takagahara00,bayer02}.
For the cavity damping rate $\kappa$,
we considered two cases:
the weak cavity damping case $\kappa\ll\Delta\omega_{\text{ph}}$,
and the strong cavity damping case $\kappa\gg\Delta\omega_{\text{ph}}$.
We also assumed the resonance condition 
$\omega_{\text{ph}}=\omega_{\text{cav}}=\omega_{\text{ex}}$, and
the input photon polarization was $\ket{\sigma^{+}}$.

%\section{Results}

%%%%%%%%%%%%%%%%%%%%%%%%%%%%%%%%%%%%%%%%%%%%%%%%%%
%%%%% result
%%%%%%%%%%%%%%%%%%%%%%%%%%%%%%%%%%%%%%%%%%%%%%%%%

%%%%% Figure yield_weakcavitydamping
\begin{figure}[t]
 \begin{center}
  \begin{tabular}{ll}
   (a) & (b)
   \\
   \includegraphics[width=0.49\linewidth,clip]{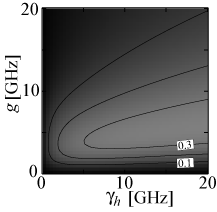}
   &
       \includegraphics[width=0.49\linewidth,clip]{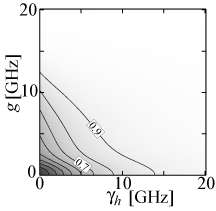}
  \end{tabular}
 \end{center}
 \caption{Contour plots of (a) yield and (b) fidelity,
 as functions of the extraction rate of hole $\gamma_h$ and
 cavity-dot coupling $g$, in the case of weak cavity damping.
 The other parameters were as follows:
 input photon bandwidth, $\Delta\omega_{\text{ph}}=5\text{GHz}$;
 cavity damping rate, $\kappa=0.5\text{GHz}$;
 electron-hole exchange interaction, $\omega_J=5\text{GHz}$;
 and spontaneous emission rate, $\gamma_{SE}=1\text{GHz}$.}
 \label{fig:yield_weakcavitydamping}
 \label{fig:fidelity_weakcavitydamping}
\end{figure}
%[NOTE. THIS FIGURE IS CURRENTLY PLACED ON THE PAGE BEFORE THE ONE ON WHICH IT IS FIRST MENTIONED. YOU MAY WISH TO MOVE IT CLOSER TO THE POINT IN THE TEXT WHERE YOU REFER TO IT]

In what follows, we will show the results for yield and fidelity
in the weak ($\kappa=0.5\text{GHz}$) 
and strong ($\kappa=15\text{GHz}$) cavity damping cases.
We will also discuss the effects of 
spontaneous emission.

%%%%% weak cavity damping case
Let us discuss the yield and fidelity of the QST
in the weak cavity damping case $\kappa\ll\Delta\omega_{\text{ph}}$.
In Fig. \ref{fig:yield_weakcavitydamping} (a),
we show the contour plot of the yield 
as a function of the extraction rate of hole $\gamma_h$ and
cavity-dot coupling $g$.
We take $\kappa=0.5\text{GHz}$.
%$\Delta\omega_{ph}=5\text{GHz}$, $\omega_J=5\text{GHz}$,
%and $\gamma_{SE}=1\text{GHz}$.
As shown in the figure, the yield is much less than the unity
because the cavity damping rate $\kappa$ is so small 
that most input photons can not enter the cavity.
Figure \ref{fig:fidelity_weakcavitydamping}(b) is the contour plot
of the fidelity as a function of $\gamma_h$ and $g$.
Note that the fidelity is suppressed in the vicinity of the origin,
where the dwell time of the hole is much longer than the
precession period of the electron spin due to the 
electron-hole exchange interaction.

\begin{figure}[t]
 \begin{center}
  \begin{tabular}{ll}
   (a) & (b)
   \\
   \includegraphics[width=0.49\linewidth,clip]{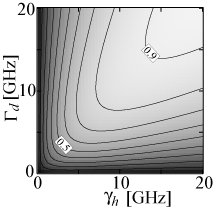}
   &
   \includegraphics[width=0.49\linewidth,clip]{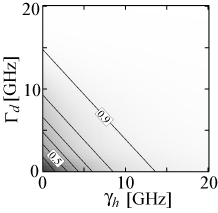}
  \end{tabular}
 \end{center}
 \caption{Contour plots of (a) yield and (b) fidelity,
 as functions of the extraction rate of hole $\gamma_h$ and
 the effective dipole relaxation rate $\Gamma_d$
 in the case of strong cavity damping.
 The other parameters were as follows:
 input photon bandwidth, $\Delta\omega_{ph}=5\text{GHz}$;
 cavity damping rate, $\kappa=15\text{GHz}$;
 electron-hole exchange interaction, $\omega_J=5\text{GHz}$;
 and spontaneous emission rate, $\gamma_{SE}=1\text{GHz}$.}
 \label{fig:yield_strongcavitydamping}
 \label{fig:fidelity_strongcavitydamping}
\end{figure}

%%%%% strong cavity damping case
In the strong cavity damping
case $\kappa\gg\Delta\omega_{\text{ph}}$, it
is useful to introduce the effective dipole relaxation rate
defined as $\Gamma_d\equiv |g|^2/\kappa$, which
represents the enhanced spontaneous emission rate of
the exciton due to the coupling with the cavity photon.
Figure \ref{fig:yield_strongcavitydamping}(a) shows the contour plot
of the yield as a function of $\Gamma_d$ and $\gamma_h$
with $\kappa=15\text{GHz}$.
To consider in detail the effects of $\gamma_h$ on the yield,
let us fix $\Gamma_d$ at a certain value and 
increase $\gamma_h$ from zero.
As $\gamma_h$ increases,
the recombination of the dot exciton is suppressed,
hence the yield increases with increasing $\gamma_h$.
However, the increase of $\gamma_h$ prevents 
the creation of the dot exciton,
which suppresses the yield for large values of $\gamma_h$.
Therefore, the yield takes its maximum value at $\gamma_h\sim\Gamma_d$.
The condition $\Delta\omega_{\text{ph}}\ll \Gamma_d, \gamma_h$
is also important for producing a high yield
in order to transfer energy effectively 
from the input photon to the exciton.
The condition for this high yield is summarized as
$\Delta\omega_{\text{ph}}\ll\gamma_h\sim\Gamma_d$.

Fig. \ref{fig:yield_strongcavitydamping}(b) is the contour plot
of the fidelity.
Note that the fidelity is reduced from unity
for $\Gamma_d+\gamma_h<\omega_J$.
The lifetime of the exciton state $\varphi^{\text{ex}}_{s}(t)$
is given as $(\Gamma_d+\gamma_h)^{-1}$ for the impulse-like 
input photon wave packet in the case of strong cavity damping.
In order to avoid the effects of the electron-hole exchange interaction,
the lifetime of the exciton state should be much shorter
than the characteristic time of the exchange interaction
$\omega_J^{-1}$.
Therefore, the condition $\omega_J\ll\Gamma_d+\gamma_h$ 
is required for a high-fidelity QST.

%%%%% Figure yield_strongcavitydamping

\begin{figure}[t]
 \begin{center}
  \begin{tabular}{ll}
   (a) & (b)
   \\
   \includegraphics[width=0.49\linewidth,clip]{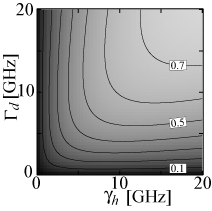}
   &
       \includegraphics[width=0.49\linewidth,clip]{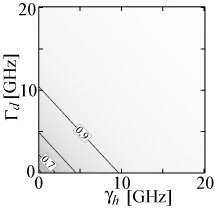}
  \end{tabular}
 \end{center}
 \caption{Contour plots of (a) yield and (b) fidelity,
 as functions of $\gamma_h$ and $\Gamma_d$
 in the case of the relatively large spontaneous emission rate 
 $\gamma_{SE}=5\text{GHz}$.
 We assume $\Delta\omega_{ph}=5\text{GHz}$, $\kappa=15\text{GHz}$, and
 $\omega_J=5\text{GHz}$.}
 \label{fig:yield_strongSE}
 \label{fig:fidelity_strongSE}
\end{figure}

%%%%%% strong SE rate
Let us now examine the effects of
spontaneous emission on the yield and fidelity.
Since we have taken the spontaneous emission rate 
$\gamma_{\text{SE}}=1\text{GHz}$ to be much smaller than the
input photon bandwidth $\Delta\omega_{\text{ph}}=5\text{GHz}$ 
in the above discussions,
the yield and fidelity 
are not significantly affected by spontaneous emission.
Figures \ref{fig:yield_strongSE}(a) and (b) show the contour plots of
yield and fidelity, respectively, in the case of a relatively 
large spontaneous emission rate $\gamma_{\text{SE}}=5\text{GHz}$.
When the spontaneous emission process is dominant,
the emission occurs before 
the hole is extracted from the dot, 
and the yield decreases as the spontaneous emission rate
$\gamma_{\text{SE}}$ increases, as shown in Fig. \ref{fig:yield_strongSE}(a).
However, even in the case of a large spontaneous emission rate,
we can obtain high yield by setting $\gamma_h$ and $\Gamma_d$ 
to be much greater than $\gamma_{\text{SE}}$, and 
by satisfying the matching condition, $\gamma_h\sim\Gamma_d$.
If we compare Figs. \ref{fig:fidelity_strongcavitydamping}(b)
and \ref{fig:fidelity_strongSE}(b),
we recognize that the fidelity is slightly improved by
taking a large $\gamma_{SE}$.
The lifetime of the exciton is shortened by the spontaneous emission,
and the spin-flip process of the electron-hole exchange interaction 
is therefore suppressed and fidelity is improved.

%%%%% Difference between previous work.
With respect to the relationship between the results obtained
in the present study and those of our previous work \cite{rikitake06}, 
we must point out that, in our previous paper, 
we analyzed the fidelity of the QST
by considering the QST process after the creation of the dot exciton,
but did not consider the yield of the QST.
We concluded in that work that the hole should be extracted from the dot
as rapidly as possible in order to achieve high fidelity,
which is consistent with the present results.
That is, we can achieve higher fidelity 
as the extraction rate of the hole increases,
as shown in Fig. \ref{fig:fidelity_strongcavitydamping} (b).
However, too large a value of $\gamma_h$ leads to a decrease in the yield, 
as shown in Fig. \ref{fig:yield_strongcavitydamping} (a).
In order to obtain both high yield and high fidelity, 
$\gamma_h$ and $g$ should be as large as possible
while satisfying the matching condition 
$\gamma_d\sim\Gamma_d\equiv |g|^2/\kappa$.

%%%%% Example estimation %%%% 
Finally, we estimated the yield and fidelity of the QST
in a single quantum dot-semiconductor microcavity coupling system
\cite{reithmaier04,yoshie04,peter06}.
%with realistic parameters in the case of strong cavity damping.
Let us consider the GaAs quantum dot formed by 
monolayer fluctuations in the
Al$_{0.8}$In$_{0.2}$As/GaAs/Al$_{0.8}$In$_{0.2}$As 
quantum well (QW).
By adjusting the thickness of the QW appropriately, 
We can set the electron spin $g$-factor in the QW to zero
\cite{nakayama92} 
and the energy levels shown in Fig. \ref{fig:SelectionRule}(a) can be realized.
The excitation energy of the dot exciton is estimated about $1.6\text{eV}$.
Note that as shown in Fig. \ref{fig:SelectionRule}(b),
the continuum of the hole state should be
attached via AlInAs barrier layer to extract the created hole.
As the cavity we consider the micropllar cavity having
top and bottom Al$_x$Ga$_{1-x}$As/AlAs ($x\sim 0.27$) 
Bragg mirrors\cite{andreani99}, %% << new ref! >>
and the QW is located in the cavity.
The cavity-dot coupling is given by 
$g=(\frac{1}{4\pi\epsilon_0\epsilon_r}\frac{\pi e^2 f}{mV})^{1/2}$,
where $f$ is the exciton oscillator strength, 
$V$ the effective modal volume, $m$ the free electron mass,
$e$ the electron charge, $\epsilon_0$ the electric constant, and
$\epsilon_r$ the relative permittivity.
The oscillator strength of transition from heavy heavy hole states
(denoted as hh$\pm$ in Fig. \ref{fig:SelectionRule}(a)) 
in a GaAs quantum dot $f_{hh}$ more than $100$
is theoretically predicted \cite{andreani99}
and experimentally observed\cite{peter06}. 
Because we use the transition from the lh+ state and 
its oscillator strength is roughly given by $f=f_{hh}/12$,
therefore we can assume $f\sim 10$ here.
With the cavity modal value $V\sim1\mu\text{m}^3$,
the cavity-dot coupling becomes $g=25\text{GHz}$.
For the cavity decay rate, we assume $\kappa=40\text{GHz}$,
and this value corresponds to the cavity Q-factor 
$Q\equiv\omega_{\text{cav}}/\kappa\sim 10^4$ which
now be realized in micropillar cavities\cite{reitzenstein06}. 
%% << new ref! >>
The extraction rate of the hole $\gamma_h$ is a tunable parameter
by changing the barrier width.
Here we take $\gamma_h=15.6\text{GHz}$ so that the
matching condition $\gamma_d\sim\Gamma_d\equiv |g|^2/\kappa$ is
satisfied.
With the other parameters,
input photon bandwidth $\Delta\omega_{\text{ph}}=5\text{GHz}$,
electron-hole exchange interaction $\omega_J=5\text{GHz}$, and
and spontaneous emission rate $\gamma_{\text{SE}}=1\text{GHz}$,
we then obtain yield $P=0.913$ and 
fidelity $F=0.976$.

%\section{Conclusion}
%%%%%%%%%%%%%%%%%%%%%%%%%%%%%%%%%%%%%%%%%%%%%%%%%%
%%%%% conclusion 
%%%%%%%%%%%%%%%%%%%%%%%%%%%%%%%%%%%%%%%%%%%%%%%%%%
In conclusion,
we have analyzed the yield and fidelity of the
QST from the photon qubit
to the spin qubit in a spin-coherent semiconductor photo detector.
By considering a model in which the quantum dot is coupled with the 
photonic cavity, we determined the optimal conditions 
for high-yield and high-fidelity QST. 
Briefly, the cavity damping rate should be larger
than the input photon bandwidth, $\Delta\omega_{\text{ph}}\ll\kappa$, 
and the matching condition, $\omega_{\text{ph}}\ll\gamma_h\sim\Gamma_d$,
between the extraction rate of the hole and 
the effective dipole relaxation rate
should be also satisfied
in order to obtain high yield.
For high-fidelity QST, a small electron-hole exchange interaction
is preferred compared to the extraction rate of the hole 
and the effective dipole relaxation rate, $\omega_J\ll\Gamma_d+\gamma_h$.

%\section*{Acknowledgment}

The authors would like to thank T. Takagahara for valuable discussions.
This work was supported 
by  CREST, MEXT. KAKENHI (No. 16710061), 
the NAREGI Nanoscience Project, and a NEDO Grant.

%%%%%%%%%%%%%%%%%%%%%%%%%%%%%%%%%%%%%%%%%%%%%%%%%%
%%%%%%%%%%%%%%%%%%%%%%%%%%%%%%%%%%%%%%%%%%%%%%%%%%
%%%%%%%%%%%%%%%%%%%%%%%%%%%%%%%%%%%%%%%%%%%%%%%%%%

%\bibliography{refs_new}

\begin{thebibliography}{40}
\expandafter\ifx\csname natexlab\endcsname\relax\def\natexlab#1{#1}\fi
\expandafter\ifx\csname bibnamefont\endcsname\relax
  \def\bibnamefont#1{#1}\fi
\expandafter\ifx\csname bibfnamefont\endcsname\relax
  \def\bibfnamefont#1{#1}\fi
\expandafter\ifx\csname citenamefont\endcsname\relax
  \def\citenamefont#1{#1}\fi
\expandafter\ifx\csname url\endcsname\relax
  \def\url#1{\texttt{#1}}\fi
\expandafter\ifx\csname urlprefix\endcsname\relax\def\urlprefix{URL }\fi
\providecommand{\bibinfo}[2]{#2}
\providecommand{\eprint}[2][]{\url{#2}}

\bibitem[{\citenamefont{Briegel et~al.}(1998)\citenamefont{Briegel, Dur, Cirac,
  and Zoller}}]{briegel98}
\bibinfo{author}{\bibfnamefont{H.}~\bibnamefont{Briegel}},
  \bibinfo{author}{\bibfnamefont{W.}~\bibnamefont{Dur}},
  \bibinfo{author}{\bibfnamefont{J.}~\bibnamefont{Cirac}}, \bibnamefont{and}
  \bibinfo{author}{\bibfnamefont{P.}~\bibnamefont{Zoller}},
  \bibinfo{journal}{Phys. Rev. Lett.} \textbf{\bibinfo{volume}{81}},
  \bibinfo{pages}{5932} (\bibinfo{year}{1998}).

\bibitem[{\citenamefont{Childress et~al.}(2005)\citenamefont{Childress, Taylor,
  Sorensen, and Lukin}}]{childress05}
\bibinfo{author}{\bibfnamefont{L.}~\bibnamefont{Childress}},
  \bibinfo{author}{\bibfnamefont{J.}~\bibnamefont{Taylor}},
  \bibinfo{author}{\bibfnamefont{A.}~\bibnamefont{Sorensen}}, \bibnamefont{and}
  \bibinfo{author}{\bibfnamefont{M.}~\bibnamefont{Lukin}},
  \bibinfo{journal}{Phys. Rev. A} \textbf{\bibinfo{volume}{72}},
  \bibinfo{pages}{052330} (\bibinfo{year}{2005}).

\bibitem[{\citenamefont{Taylor et~al.}(2005)\citenamefont{Taylor, Dur, Zoller,
  Yacoby, Marcus, and Lukin}}]{taylor05}
\bibinfo{author}{\bibfnamefont{J.}~\bibnamefont{Taylor}},
  \bibinfo{author}{\bibfnamefont{W.}~\bibnamefont{Dur}},
  \bibinfo{author}{\bibfnamefont{P.}~\bibnamefont{Zoller}},
  \bibinfo{author}{\bibfnamefont{A.}~\bibnamefont{Yacoby}},
  \bibinfo{author}{\bibfnamefont{C.}~\bibnamefont{Marcus}}, \bibnamefont{and}
  \bibinfo{author}{\bibfnamefont{M.}~\bibnamefont{Lukin}},
  \bibinfo{journal}{Phys. Rev. Lett.} \textbf{\bibinfo{volume}{94}},
  \bibinfo{pages}{236803} (\bibinfo{year}{2005}).

\bibitem[{\citenamefont{Gisin et~al.}(2002)\citenamefont{Gisin, Ribordy,
  Tittel, and Zbinden}}]{gisin02}
\bibinfo{author}{\bibfnamefont{N.}~\bibnamefont{Gisin}},
  \bibinfo{author}{\bibfnamefont{G.}~\bibnamefont{Ribordy}},
  \bibinfo{author}{\bibfnamefont{W.}~\bibnamefont{Tittel}}, \bibnamefont{and}
  \bibinfo{author}{\bibfnamefont{H.}~\bibnamefont{Zbinden}},
  \bibinfo{journal}{Rev. Mod. Phys.} \textbf{\bibinfo{volume}{74}},
  \bibinfo{pages}{145} (\bibinfo{year}{2002}).

\bibitem[{\citenamefont{Gobby et~al.}(2004)\citenamefont{Gobby, Yuan, and
  Shields}}]{gobby04}
\bibinfo{author}{\bibfnamefont{C.}~\bibnamefont{Gobby}},
  \bibinfo{author}{\bibfnamefont{Z.~L.} \bibnamefont{Yuan}}, \bibnamefont{and}
  \bibinfo{author}{\bibfnamefont{A.~J.} \bibnamefont{Shields}},
  \bibinfo{journal}{Appl. Phys. Lett.} \textbf{\bibinfo{volume}{84}},
  \bibinfo{pages}{3762} (\bibinfo{year}{2004}).

\bibitem[{\citenamefont{Loss and DiVincenzo}(1998)}]{loss98}
\bibinfo{author}{\bibfnamefont{D.}~\bibnamefont{Loss}} \bibnamefont{and}
  \bibinfo{author}{\bibfnamefont{D.}~\bibnamefont{DiVincenzo}},
  \bibinfo{journal}{Phys. Rev. A} \textbf{\bibinfo{volume}{57}},
  \bibinfo{pages}{120} (\bibinfo{year}{1998}).

\bibitem[{\citenamefont{Taylor et~al.}(2003)\citenamefont{Taylor, Marcus, and
  Lukin}}]{taylor03}
\bibinfo{author}{\bibfnamefont{J.}~\bibnamefont{Taylor}},
  \bibinfo{author}{\bibfnamefont{C.}~\bibnamefont{Marcus}}, \bibnamefont{and}
  \bibinfo{author}{\bibfnamefont{M.}~\bibnamefont{Lukin}},
  \bibinfo{journal}{Phys. Rev. Lett.} \textbf{\bibinfo{volume}{90}},
  \bibinfo{pages}{206803} (\bibinfo{year}{2003}).

\bibitem[{\citenamefont{Petta et~al.}(2005{\natexlab{a}})\citenamefont{Petta,
  Johnson, Taylor, Laird, Yacoby, Lukin, Marcus, Hanson, and
  Gossord}}]{petta05}
\bibinfo{author}{\bibfnamefont{J.~R.} \bibnamefont{Petta}},
  \bibinfo{author}{\bibfnamefont{A.~C.} \bibnamefont{Johnson}},
  \bibinfo{author}{\bibfnamefont{J.~M.} \bibnamefont{Taylor}},
  \bibinfo{author}{\bibfnamefont{E.~A.} \bibnamefont{Laird}},
  \bibinfo{author}{\bibfnamefont{A.}~\bibnamefont{Yacoby}},
  \bibinfo{author}{\bibfnamefont{M.~D.} \bibnamefont{Lukin}},
  \bibinfo{author}{\bibfnamefont{C.~M.} \bibnamefont{Marcus}},
  \bibinfo{author}{\bibfnamefont{M.~P.} \bibnamefont{Hanson}},
  \bibnamefont{and} \bibinfo{author}{\bibfnamefont{A.~C.}
  \bibnamefont{Gossord}}, \bibinfo{journal}{Science}
  \textbf{\bibinfo{volume}{309}}, \bibinfo{pages}{2180}
  (\bibinfo{year}{2005}{\natexlab{a}}).

\bibitem[{\citenamefont{Petta et~al.}(2005{\natexlab{b}})\citenamefont{Petta,
  Johnson, Yacoby, Marcus, Hanson, and Gossard}}]{petta05a}
\bibinfo{author}{\bibfnamefont{J.}~\bibnamefont{Petta}},
  \bibinfo{author}{\bibfnamefont{A.}~\bibnamefont{Johnson}},
  \bibinfo{author}{\bibfnamefont{A.}~\bibnamefont{Yacoby}},
  \bibinfo{author}{\bibfnamefont{C.}~\bibnamefont{Marcus}},
  \bibinfo{author}{\bibfnamefont{M.}~\bibnamefont{Hanson}}, \bibnamefont{and}
  \bibinfo{author}{\bibfnamefont{A.}~\bibnamefont{Gossard}},
  \bibinfo{journal}{Phys. Rev. B} \textbf{\bibinfo{volume}{72}},
  \bibinfo{pages}{161301} (\bibinfo{year}{2005}{\natexlab{b}}).

\bibitem[{\citenamefont{Vrijen and Yablonovitch}(2001)}]{vrijen01}
\bibinfo{author}{\bibfnamefont{R.}~\bibnamefont{Vrijen}} \bibnamefont{and}
  \bibinfo{author}{\bibfnamefont{E.}~\bibnamefont{Yablonovitch}},
  \bibinfo{journal}{Physica E} \textbf{\bibinfo{volume}{10}},
  \bibinfo{pages}{569} (\bibinfo{year}{2001}).

\bibitem[{\citenamefont{Duan et~al.}(2001)\citenamefont{Duan, Lukin, Cirac, and
  Zoller}}]{duan01}
\bibinfo{author}{\bibfnamefont{L.}~\bibnamefont{Duan}},
  \bibinfo{author}{\bibfnamefont{M.}~\bibnamefont{Lukin}},
  \bibinfo{author}{\bibfnamefont{J.}~\bibnamefont{Cirac}}, \bibnamefont{and}
  \bibinfo{author}{\bibfnamefont{P.}~\bibnamefont{Zoller}},
  \bibinfo{journal}{Nature} \textbf{\bibinfo{volume}{414}},
  \bibinfo{pages}{413} (\bibinfo{year}{2001}).

\bibitem[{\citenamefont{Kosaka et~al.}(2003)\citenamefont{Kosaka, Rao,
  Robinson, Bandaru, Makita, and Yablonovitch}}]{kosaka03}
\bibinfo{author}{\bibfnamefont{H.}~\bibnamefont{Kosaka}},
  \bibinfo{author}{\bibfnamefont{D.}~\bibnamefont{Rao}},
  \bibinfo{author}{\bibfnamefont{H.}~\bibnamefont{Robinson}},
  \bibinfo{author}{\bibfnamefont{P.}~\bibnamefont{Bandaru}},
  \bibinfo{author}{\bibfnamefont{K.}~\bibnamefont{Makita}}, \bibnamefont{and}
  \bibinfo{author}{\bibfnamefont{E.}~\bibnamefont{Yablonovitch}},
  \bibinfo{journal}{Phys. Rev. B} \textbf{\bibinfo{volume}{67}},
  \bibinfo{pages}{045104} (\bibinfo{year}{2003}).

\bibitem[{\citenamefont{Muto et~al.}(2005)\citenamefont{Muto, Adachi, Yokoi,
  Sasakura, and Suemune}}]{muto05}
\bibinfo{author}{\bibfnamefont{S.}~\bibnamefont{Muto}},
  \bibinfo{author}{\bibfnamefont{S.}~\bibnamefont{Adachi}},
  \bibinfo{author}{\bibfnamefont{T.}~\bibnamefont{Yokoi}},
  \bibinfo{author}{\bibfnamefont{H.}~\bibnamefont{Sasakura}}, \bibnamefont{and}
  \bibinfo{author}{\bibfnamefont{I.}~\bibnamefont{Suemune}},
  \bibinfo{journal}{Appl. Phys. Lett.} \textbf{\bibinfo{volume}{87}},
  \bibinfo{pages}{112506} (\bibinfo{year}{2005}).

\bibitem[{\citenamefont{Kiselev et~al.}(1998)\citenamefont{Kiselev, Ivchenko,
  and Rossler}}]{kiselev98}
\bibinfo{author}{\bibfnamefont{A.}~\bibnamefont{Kiselev}},
  \bibinfo{author}{\bibfnamefont{E.}~\bibnamefont{Ivchenko}}, \bibnamefont{and}
  \bibinfo{author}{\bibfnamefont{U.}~\bibnamefont{Rossler}},
  \bibinfo{journal}{Phys. Rev. B} \textbf{\bibinfo{volume}{58}},
  \bibinfo{pages}{16353} (\bibinfo{year}{1998}).

\bibitem[{\citenamefont{Ivchenko and Kiselev}(1998)}]{ivchenko98}
\bibinfo{author}{\bibfnamefont{E.}~\bibnamefont{Ivchenko}} \bibnamefont{and}
  \bibinfo{author}{\bibfnamefont{A.}~\bibnamefont{Kiselev}},
  \bibinfo{journal}{Jetp Lett.} \textbf{\bibinfo{volume}{67}},
  \bibinfo{pages}{43} (\bibinfo{year}{1998}).

\bibitem[{\citenamefont{Matveev et~al.}(2000)\citenamefont{Matveev, Glazman,
  and Larkin}}]{matveev00}
\bibinfo{author}{\bibfnamefont{K.}~\bibnamefont{Matveev}},
  \bibinfo{author}{\bibfnamefont{L.}~\bibnamefont{Glazman}}, \bibnamefont{and}
  \bibinfo{author}{\bibfnamefont{A.}~\bibnamefont{Larkin}},
  \bibinfo{journal}{Phys. Rev. Lett.} \textbf{\bibinfo{volume}{85}},
  \bibinfo{pages}{2789} (\bibinfo{year}{2000}).

\bibitem[{\citenamefont{Kosaka et~al.}(2001)\citenamefont{Kosaka, Kiselev,
  Baron, Kim, and Yablonovitch}}]{kosaka01}
\bibinfo{author}{\bibfnamefont{H.}~\bibnamefont{Kosaka}},
  \bibinfo{author}{\bibfnamefont{A.}~\bibnamefont{Kiselev}},
  \bibinfo{author}{\bibfnamefont{F.}~\bibnamefont{Baron}},
  \bibinfo{author}{\bibfnamefont{K.}~\bibnamefont{Kim}}, \bibnamefont{and}
  \bibinfo{author}{\bibfnamefont{E.}~\bibnamefont{Yablonovitch}},
  \bibinfo{journal}{Electron. Lett.} \textbf{\bibinfo{volume}{37}},
  \bibinfo{pages}{464} (\bibinfo{year}{2001}).

\bibitem[{\citenamefont{Salis et~al.}(2001)\citenamefont{Salis, Kato, Ensslin,
  Driscoll, Gossard, and Awschalom}}]{salis01}
\bibinfo{author}{\bibfnamefont{G.}~\bibnamefont{Salis}},
  \bibinfo{author}{\bibfnamefont{Y.}~\bibnamefont{Kato}},
  \bibinfo{author}{\bibfnamefont{K.}~\bibnamefont{Ensslin}},
  \bibinfo{author}{\bibfnamefont{D.}~\bibnamefont{Driscoll}},
  \bibinfo{author}{\bibfnamefont{A.}~\bibnamefont{Gossard}}, \bibnamefont{and}
  \bibinfo{author}{\bibfnamefont{D.}~\bibnamefont{Awschalom}},
  \bibinfo{journal}{Nature} \textbf{\bibinfo{volume}{414}},
  \bibinfo{pages}{619} (\bibinfo{year}{2001}).

\bibitem[{\citenamefont{Salis et~al.}(2003)\citenamefont{Salis, Kato, Ensslin,
  Driscoll, Gossard, and Awschalom}}]{salis03}
\bibinfo{author}{\bibfnamefont{G.}~\bibnamefont{Salis}},
  \bibinfo{author}{\bibfnamefont{Y.}~\bibnamefont{Kato}},
  \bibinfo{author}{\bibfnamefont{K.}~\bibnamefont{Ensslin}},
  \bibinfo{author}{\bibfnamefont{D.}~\bibnamefont{Driscoll}},
  \bibinfo{author}{\bibfnamefont{A.}~\bibnamefont{Gossard}}, \bibnamefont{and}
  \bibinfo{author}{\bibfnamefont{D.}~\bibnamefont{Awschalom}},
  \bibinfo{journal}{Physica E} \textbf{\bibinfo{volume}{16}},
  \bibinfo{pages}{99} (\bibinfo{year}{2003}).

\bibitem[{\citenamefont{Nitta et~al.}(2003)\citenamefont{Nitta, Lin, Akazaki,
  and Koga}}]{nitta03}
\bibinfo{author}{\bibfnamefont{J.}~\bibnamefont{Nitta}},
  \bibinfo{author}{\bibfnamefont{Y.}~\bibnamefont{Lin}},
  \bibinfo{author}{\bibfnamefont{T.}~\bibnamefont{Akazaki}}, \bibnamefont{and}
  \bibinfo{author}{\bibfnamefont{T.}~\bibnamefont{Koga}},
  \bibinfo{journal}{Appl. Phys. Lett.} \textbf{\bibinfo{volume}{83}},
  \bibinfo{pages}{4565} (\bibinfo{year}{2003}).

\bibitem[{\citenamefont{Lin et~al.}(2004)\citenamefont{Lin, Nitta, Koga, and
  Akazaki}}]{lin04}
\bibinfo{author}{\bibfnamefont{Y.}~\bibnamefont{Lin}},
  \bibinfo{author}{\bibfnamefont{J.}~\bibnamefont{Nitta}},
  \bibinfo{author}{\bibfnamefont{T.}~\bibnamefont{Koga}}, \bibnamefont{and}
  \bibinfo{author}{\bibfnamefont{T.}~\bibnamefont{Akazaki}},
  \bibinfo{journal}{Physica E} \textbf{\bibinfo{volume}{21}},
  \bibinfo{pages}{656} (\bibinfo{year}{2004}).

\bibitem[{\citenamefont{Nitta et~al.}(2004)\citenamefont{Nitta, Lin, Koga, and
  Akazaki}}]{nitta04}
\bibinfo{author}{\bibfnamefont{J.}~\bibnamefont{Nitta}},
  \bibinfo{author}{\bibfnamefont{Y.}~\bibnamefont{Lin}},
  \bibinfo{author}{\bibfnamefont{T.}~\bibnamefont{Koga}}, \bibnamefont{and}
  \bibinfo{author}{\bibfnamefont{T.}~\bibnamefont{Akazaki}},
  \bibinfo{journal}{Physica E} \textbf{\bibinfo{volume}{20}},
  \bibinfo{pages}{429} (\bibinfo{year}{2004}).

\bibitem[{\citenamefont{Lin et~al.}(1991)\citenamefont{Lin, Wei, Tsui, Klem,
  and Allen}}]{lin91}
\bibinfo{author}{\bibfnamefont{S.}~\bibnamefont{Lin}},
  \bibinfo{author}{\bibfnamefont{H.}~\bibnamefont{Wei}},
  \bibinfo{author}{\bibfnamefont{D.}~\bibnamefont{Tsui}},
  \bibinfo{author}{\bibfnamefont{J.}~\bibnamefont{Klem}}, \bibnamefont{and}
  \bibinfo{author}{\bibfnamefont{S.}~\bibnamefont{Allen}},
  \bibinfo{journal}{Phys. Rev. B} \textbf{\bibinfo{volume}{43}},
  \bibinfo{pages}{12110} (\bibinfo{year}{1991}).

\bibitem[{\citenamefont{Nakaoka et~al.}(2004)\citenamefont{Nakaoka, Saito,
  Tatebayashi, and Arakawa}}]{nakaoka04}
\bibinfo{author}{\bibfnamefont{T.}~\bibnamefont{Nakaoka}},
  \bibinfo{author}{\bibfnamefont{T.}~\bibnamefont{Saito}},
  \bibinfo{author}{\bibfnamefont{J.}~\bibnamefont{Tatebayashi}},
  \bibnamefont{and} \bibinfo{author}{\bibfnamefont{Y.}~\bibnamefont{Arakawa}},
  \bibinfo{journal}{Phys. Rev. B} \textbf{\bibinfo{volume}{70}},
  \bibinfo{pages}{235337} (\bibinfo{year}{2004}).

\bibitem[{\citenamefont{Nakaoka et~al.}(2005)\citenamefont{Nakaoka, Saito,
  Tatebayashi, Hirose, Usuki, Yokoyama, and Arakawa}}]{nakaoka05}
\bibinfo{author}{\bibfnamefont{T.}~\bibnamefont{Nakaoka}},
  \bibinfo{author}{\bibfnamefont{T.}~\bibnamefont{Saito}},
  \bibinfo{author}{\bibfnamefont{J.}~\bibnamefont{Tatebayashi}},
  \bibinfo{author}{\bibfnamefont{S.}~\bibnamefont{Hirose}},
  \bibinfo{author}{\bibfnamefont{T.}~\bibnamefont{Usuki}},
  \bibinfo{author}{\bibfnamefont{N.}~\bibnamefont{Yokoyama}}, \bibnamefont{and}
  \bibinfo{author}{\bibfnamefont{Y.}~\bibnamefont{Arakawa}},
  \bibinfo{journal}{Phys. Rev. B} \textbf{\bibinfo{volume}{71}},
  \bibinfo{pages}{205301} (\bibinfo{year}{2005}).

\bibitem[{\citenamefont{Rikitake and Imamura}(2006)}]{rikitake06}
\bibinfo{author}{\bibfnamefont{Y.}~\bibnamefont{Rikitake}} \bibnamefont{and}
  \bibinfo{author}{\bibfnamefont{H.}~\bibnamefont{Imamura}},
  \bibinfo{journal}{Phys. Rev. B} \textbf{\bibinfo{volume}{74}},
  \bibinfo{pages}{081307} (\bibinfo{year}{2006}).

\bibitem[{\citenamefont{Gurvitz et~al.}(1991)\citenamefont{Gurvitz, Barjoseph,
  and Deveaud}}]{gurvitz91}
\bibinfo{author}{\bibfnamefont{S.}~\bibnamefont{Gurvitz}},
  \bibinfo{author}{\bibfnamefont{I.}~\bibnamefont{Barjoseph}},
  \bibnamefont{and} \bibinfo{author}{\bibfnamefont{B.}~\bibnamefont{Deveaud}},
  \bibinfo{journal}{Phys. Rev. B} \textbf{\bibinfo{volume}{43}},
  \bibinfo{pages}{14703} (\bibinfo{year}{1991}).

\bibitem[{\citenamefont{Cohen et~al.}(1993)\citenamefont{Cohen, Gurvitz,
  Barjoseph, Deveaud, Bergman, and Regreny}}]{cohen93}
\bibinfo{author}{\bibfnamefont{G.}~\bibnamefont{Cohen}},
  \bibinfo{author}{\bibfnamefont{S.}~\bibnamefont{Gurvitz}},
  \bibinfo{author}{\bibfnamefont{I.}~\bibnamefont{Barjoseph}},
  \bibinfo{author}{\bibfnamefont{B.}~\bibnamefont{Deveaud}},
  \bibinfo{author}{\bibfnamefont{P.}~\bibnamefont{Bergman}}, \bibnamefont{and}
  \bibinfo{author}{\bibfnamefont{A.}~\bibnamefont{Regreny}},
  \bibinfo{journal}{Phys. Rev. B} \textbf{\bibinfo{volume}{47}},
  \bibinfo{pages}{16012} (\bibinfo{year}{1993}).

\bibitem[{\citenamefont{Kojima et~al.}(2003)\citenamefont{Kojima, Hofmann,
  Takeuchi, and Sasaki}}]{kojima03}
\bibinfo{author}{\bibfnamefont{K.}~\bibnamefont{Kojima}},
  \bibinfo{author}{\bibfnamefont{H.}~\bibnamefont{Hofmann}},
  \bibinfo{author}{\bibfnamefont{S.}~\bibnamefont{Takeuchi}}, \bibnamefont{and}
  \bibinfo{author}{\bibfnamefont{K.}~\bibnamefont{Sasaki}},
  \bibinfo{journal}{Phys. Rev. A} \textbf{\bibinfo{volume}{68}},
  \bibinfo{pages}{013803} (\bibinfo{year}{2003}).

\bibitem[{\citenamefont{Kojima and Tomita}(2007)}]{kojima07}
\bibinfo{author}{\bibfnamefont{K.}~\bibnamefont{Kojima}} \bibnamefont{and}
  \bibinfo{author}{\bibfnamefont{A.}~\bibnamefont{Tomita}},
  \bibinfo{journal}{Phys. Rev. A} \textbf{\bibinfo{volume}{75}}
  (\bibinfo{year}{2007}).

\bibitem[{\citenamefont{Koshino and Ishihara}(2004)}]{koshino04b}
\bibinfo{author}{\bibfnamefont{K.}~\bibnamefont{Koshino}} \bibnamefont{and}
  \bibinfo{author}{\bibfnamefont{H.}~\bibnamefont{Ishihara}},
  \bibinfo{journal}{Phys. Rev. A} \textbf{\bibinfo{volume}{70}},
  \bibinfo{pages}{013806} (\bibinfo{year}{2004}).

\bibitem[{\citenamefont{Gammon et~al.}(1996)\citenamefont{Gammon, Snow,
  Shanabrook, Katzer, and Park}}]{gammon96}
\bibinfo{author}{\bibfnamefont{D.}~\bibnamefont{Gammon}},
  \bibinfo{author}{\bibfnamefont{E.}~\bibnamefont{Snow}},
  \bibinfo{author}{\bibfnamefont{B.}~\bibnamefont{Shanabrook}},
  \bibinfo{author}{\bibfnamefont{D.}~\bibnamefont{Katzer}}, \bibnamefont{and}
  \bibinfo{author}{\bibfnamefont{D.}~\bibnamefont{Park}},
  \bibinfo{journal}{Phys. Rev. Lett.} \textbf{\bibinfo{volume}{76}},
  \bibinfo{pages}{3005} (\bibinfo{year}{1996}).

\bibitem[{\citenamefont{Takagahara}(2000)}]{takagahara00}
\bibinfo{author}{\bibfnamefont{T.}~\bibnamefont{Takagahara}},
  \bibinfo{journal}{Phys. Rev. B} \textbf{\bibinfo{volume}{62}},
  \bibinfo{pages}{16840} (\bibinfo{year}{2000}).

\bibitem[{\citenamefont{Bayer et~al.}(2002)\citenamefont{Bayer, Ortner, Stern,
  Kuther, Gorbunov, Forchel, Hawrylak, Fafard, Hinzer, Reinecke
  et~al.}}]{bayer02}
\bibinfo{author}{\bibfnamefont{M.}~\bibnamefont{Bayer}},
  \bibinfo{author}{\bibfnamefont{G.}~\bibnamefont{Ortner}},
  \bibinfo{author}{\bibfnamefont{O.}~\bibnamefont{Stern}},
  \bibinfo{author}{\bibfnamefont{A.}~\bibnamefont{Kuther}},
  \bibinfo{author}{\bibfnamefont{A.}~\bibnamefont{Gorbunov}},
  \bibinfo{author}{\bibfnamefont{A.}~\bibnamefont{Forchel}},
  \bibinfo{author}{\bibfnamefont{P.}~\bibnamefont{Hawrylak}},
  \bibinfo{author}{\bibfnamefont{S.}~\bibnamefont{Fafard}},
  \bibinfo{author}{\bibfnamefont{K.}~\bibnamefont{Hinzer}},
  \bibinfo{author}{\bibfnamefont{T.}~\bibnamefont{Reinecke}},
  \bibnamefont{et~al.}, \bibinfo{journal}{Phys. Rev. B}
  \textbf{\bibinfo{volume}{65}}, \bibinfo{pages}{195315}
  (\bibinfo{year}{2002}).

\bibitem[{\citenamefont{Reithmaier et~al.}(2004)\citenamefont{Reithmaier, Sek,
  Loffler, Hofmann, Kuhn, Reitzenstein, Keldysh, Kulakovskii, Reinecke, and
  Forchel}}]{reithmaier04}
\bibinfo{author}{\bibfnamefont{J.~P.} \bibnamefont{Reithmaier}},
  \bibinfo{author}{\bibfnamefont{G.}~\bibnamefont{Sek}},
  \bibinfo{author}{\bibfnamefont{A.}~\bibnamefont{Loffler}},
  \bibinfo{author}{\bibfnamefont{C.}~\bibnamefont{Hofmann}},
  \bibinfo{author}{\bibfnamefont{S.}~\bibnamefont{Kuhn}},
  \bibinfo{author}{\bibfnamefont{S.}~\bibnamefont{Reitzenstein}},
  \bibinfo{author}{\bibfnamefont{L.~V.} \bibnamefont{Keldysh}},
  \bibinfo{author}{\bibfnamefont{V.~D.} \bibnamefont{Kulakovskii}},
  \bibinfo{author}{\bibfnamefont{T.~L.} \bibnamefont{Reinecke}},
  \bibnamefont{and} \bibinfo{author}{\bibfnamefont{A.}~\bibnamefont{Forchel}},
  \bibinfo{journal}{Nature} \textbf{\bibinfo{volume}{432}},
  \bibinfo{pages}{197} (\bibinfo{year}{2004}).

\bibitem[{\citenamefont{Yoshie et~al.}(2004)\citenamefont{Yoshie, Scherer,
  Hendrickson, Khitrova, Gibbs, Rupper, Ell, Shchekin, and Deppe}}]{yoshie04}
\bibinfo{author}{\bibfnamefont{T.}~\bibnamefont{Yoshie}},
  \bibinfo{author}{\bibfnamefont{A.}~\bibnamefont{Scherer}},
  \bibinfo{author}{\bibfnamefont{J.}~\bibnamefont{Hendrickson}},
  \bibinfo{author}{\bibfnamefont{G.}~\bibnamefont{Khitrova}},
  \bibinfo{author}{\bibfnamefont{H.~M.} \bibnamefont{Gibbs}},
  \bibinfo{author}{\bibfnamefont{G.}~\bibnamefont{Rupper}},
  \bibinfo{author}{\bibfnamefont{C.}~\bibnamefont{Ell}},
  \bibinfo{author}{\bibfnamefont{O.~B.} \bibnamefont{Shchekin}},
  \bibnamefont{and} \bibinfo{author}{\bibfnamefont{D.~G.} \bibnamefont{Deppe}},
  \bibinfo{journal}{Nature} \textbf{\bibinfo{volume}{432}},
  \bibinfo{pages}{200} (\bibinfo{year}{2004}).

\bibitem[{\citenamefont{Peter et~al.}(2006)\citenamefont{Peter, Bloch, Martrou,
  Lemaitre, Hours, Patriarche, Cavanna, Gerard, Laurent, Robert-Philip
  et~al.}}]{peter06}
\bibinfo{author}{\bibfnamefont{E.}~\bibnamefont{Peter}},
  \bibinfo{author}{\bibfnamefont{J.}~\bibnamefont{Bloch}},
  \bibinfo{author}{\bibfnamefont{D.}~\bibnamefont{Martrou}},
  \bibinfo{author}{\bibfnamefont{A.}~\bibnamefont{Lemaitre}},
  \bibinfo{author}{\bibfnamefont{J.}~\bibnamefont{Hours}},
  \bibinfo{author}{\bibfnamefont{G.}~\bibnamefont{Patriarche}},
  \bibinfo{author}{\bibfnamefont{A.}~\bibnamefont{Cavanna}},
  \bibinfo{author}{\bibfnamefont{J.~M.} \bibnamefont{Gerard}},
  \bibinfo{author}{\bibfnamefont{S.}~\bibnamefont{Laurent}},
  \bibinfo{author}{\bibfnamefont{I.}~\bibnamefont{Robert-Philip}},
  \bibnamefont{et~al.}, \bibinfo{journal}{Phys. Status Solidi B-Basic Solid
  State Phys.} \textbf{\bibinfo{volume}{243}}, \bibinfo{pages}{3879}
  (\bibinfo{year}{2006}).

\bibitem[{\citenamefont{Nakayama et~al.}(1992)\citenamefont{Nakayama, Doguchi,
  and Nishimura}}]{nakayama92}
\bibinfo{author}{\bibfnamefont{M.}~\bibnamefont{Nakayama}},
  \bibinfo{author}{\bibfnamefont{T.}~\bibnamefont{Doguchi}}, \bibnamefont{and}
  \bibinfo{author}{\bibfnamefont{H.}~\bibnamefont{Nishimura}},
  \bibinfo{journal}{J. Appl. Phys.} \textbf{\bibinfo{volume}{72}},
  \bibinfo{pages}{2372} (\bibinfo{year}{1992}).

\bibitem[{\citenamefont{Andreani et~al.}(1999)\citenamefont{Andreani,
  Panzarini, and Gerard}}]{andreani99}
\bibinfo{author}{\bibfnamefont{L.~C.} \bibnamefont{Andreani}},
  \bibinfo{author}{\bibfnamefont{G.}~\bibnamefont{Panzarini}},
  \bibnamefont{and} \bibinfo{author}{\bibfnamefont{J.~M.}
  \bibnamefont{Gerard}}, \bibinfo{journal}{Phys. Rev. B}
  \textbf{\bibinfo{volume}{60}}, \bibinfo{pages}{13276} (\bibinfo{year}{1999}).

\bibitem[{\citenamefont{Reitzenstein et~al.}(2006)\citenamefont{Reitzenstein,
  Hofmann, Loffler, Kubanek, Reithmaier, Kamp, Kulakovskii, Keldysh, Reinecke,
  and Forchel}}]{reitzenstein06}
\bibinfo{author}{\bibfnamefont{S.}~\bibnamefont{Reitzenstein}},
  \bibinfo{author}{\bibfnamefont{C.}~\bibnamefont{Hofmann}},
  \bibinfo{author}{\bibfnamefont{A.}~\bibnamefont{Loffler}},
  \bibinfo{author}{\bibfnamefont{A.}~\bibnamefont{Kubanek}},
  \bibinfo{author}{\bibfnamefont{J.~P.} \bibnamefont{Reithmaier}},
  \bibinfo{author}{\bibfnamefont{M.}~\bibnamefont{Kamp}},
  \bibinfo{author}{\bibfnamefont{V.~D.} \bibnamefont{Kulakovskii}},
  \bibinfo{author}{\bibfnamefont{L.~V.} \bibnamefont{Keldysh}},
  \bibinfo{author}{\bibfnamefont{T.~L.} \bibnamefont{Reinecke}},
  \bibnamefont{and} \bibinfo{author}{\bibfnamefont{A.}~\bibnamefont{Forchel}},
  \bibinfo{journal}{Phys. Status Solidi B-Basic Solid State Phys.}
  \textbf{\bibinfo{volume}{243}}, \bibinfo{pages}{2224} (\bibinfo{year}{2006}).

\end{thebibliography}

\end{document}